\documentclass[aps,prb,doublespace,twocolumn,showpacs]{revtex4}
\usepackage{amssymb}
\usepackage{amsmath}
\usepackage{graphicx}
\begin{document}
\flushbottom

\title{Electromagnetic response of layered superconductors with
broken lattice inversion symmetry}

\author{B.~Uchoa$^1$, A.~H.~Castro Neto$^2$, and G.~G. Cabrera$^1$}

\affiliation{$^1$ Instituto de F\'{\i}sica ``Gleb Wataghin''
Universidade Estadual de Campinas (UNICAMP),\\
C. P. 6165, Campinas, SP 13083-970, Brazil \\
$^2$ Department of Physics, Boston University, 590 Commonwealth Ave.,
 Boston, MA 02215}

\date{ July 17, 2003}

\begin{abstract}
We investigate the macroscopic effects of charge density waves (CDW) and 
superconductivity in layered superconducting systems with broken lattice
inversion symmetry (allowing for piezoelectricity) 
such as two dimensional (2D) transition metal dichalcogenides (TMD).  
We work with the low temperature time dependent Ginzburg-Landau theory
and study the coupling of lattice distortions and low energy CDW collective modes to
the superconducting order parameter in the presence of electromagnetic
fields. We show that superconductivity and
piezoelectricity can coexist in these singular metals. Furthermore, our study indicates 
the nature
of the quantum phase transition between a commensurate CDW phase
and the stripe phase that has been observed as a function of applied pressure.  
\end{abstract}

\pacs{74.70.Ad, 71.10.Hf, 71.45.Lr}

\maketitle

\section{Introduction}

The quasi-two-dimensional (2D) transition metal  dichalcogenides (TMD) \cite{Withers} 
2H-$\mathrm{NbS_2}$, 2H-$\mathrm{NbSe_2}$, 2H-$\mathrm{TaS_2}$ and 2H-$\mathrm{TaSe_2}$
are layered compounds where superconductivity coexists with a charge density wave (CDW) 
state \cite{Wilson}. It has been documented experimentally that these
materials present anomalous 
properties such as a decrease of the resistivity and 
decoupling between planar and c-axis transport in the CDW phase \cite{transport}, 
anomalous impurity effects on the superconductivity and non-linear Hall
effect \cite{hall}, stripe phases \cite{stripes}, and more recently 
angle resolved photoemission experiments (ARPES) have shown a quasi-particle lifetime 
diverging linearly with the energy close to the Fermi surface \cite{valla}
in contrast to ordinary Fermi liquids where the lifetime diverges quadratically
with energy. Some of these properties are similar to the ones observed in high
temperature superconductors (HTc) \cite{valla_htc}.  Contrary to HTc,
however, TMD are extremely clean systems and therefore the anomalous
behavior is undoubtedly intrinsic and sample independent. Therefore
the understanding of the anomalous metallic behavior in these 2D materials may provide
clues for the understanding of the anomalous physics that occurs in a broad class of
2D systems where superconductivity occurs.

One of us (AHCN), has proposed recently an unified picture
of the CDW and superconducting transitions that can explain some of the
anomalies observed experimentally \cite{Neto}. The theory proposes that
the mechanism of superconductivity is related to the pairing  of the
CDW elementary excitations mediated by acoustic phonons via a piezoelectric coupling.
These excitations are Dirac electrons situated in the nodes of a CDW gap, 
resembling the Fermi surface zone of graphite. In TMD, unlike the case of graphite,
the lattice inversion symmetry is broken in the CDW phase. 
The breaking of the inversion symmetry in TMD,
as documented in neutron scattering experiments \cite{neutrons}, allows for
a piezoelectric coupling between charge carriers and phonons \cite{piezo}.
Since CDW formation is usually related to nested Fermi surfaces in 
1D systems, weak 2D nesting combined with strong variations in the electron
phonon coupling due to the tight-binding nature of the electronic orbitals, 
can be responsible for the  origin of the nodal CDW order parameter. 
This model is able to correctly explain some
of the anomalous properties of TMD such as the
energy dependence of the quasi-particle lifetime (given by the imaginary part of the 
Dirac fermion self-energy), the critical dependence of the
superconducting transition with the lattice parameters, 
the reduction of the resistivity, and the metallic behavior in the CDW phase.

However, a criticism has been raised to this theory based on the fact
that piezoelectricity normally occurs in insulating systems since metals
can screen the internal electric fields \cite{varma}. However,
Dirac fermions in 2D have a vanishing density of states at the Fermi energy and
therefore do not screen electric fields \cite{mele} allowing for the existence
of a metallic state (that is, a state with gapless fermionic excitations) and
piezoelectricity. Nevertheless, the possibility of coexistence of piezoelectricity
and superconductivity is indeed surprising. In such a system it would be
possible to generate super-currents by simply squeezing the sample. 
In this work we show that there is no contradiction
between the existence of piezoelectricity and superconductivity in materials where
the low-lying excitations are Dirac fermions. 
We investigate the current and charge fluctuations that would arise in these
systems under the view point of the collective modes. 

Using path integrals we derive a
semi-classical action that describes the coupling between plasmons (responsible for screening) 
and acoustic phonons to the superconducting order parameter via the piezoelectric
coupling. Various well-known regimes are described by this action: (1) in the case of normal
electrons without superconductivity we recover the well-known results for collective modes and 
screening in 2D and 3D \cite{fetter}; (2) in the case of Dirac fermions without superconductivity
we re-obtain the results for collective modes and screening in semi-metals like graphite \cite{Shung};
(3) in the absence of piezoelectric coupling, we recover the behavior of an ordinary type II superconductor 
\cite{tinkham}. However, when we allow the piezoelectric coupling with Dirac fermions and
superconductivity, new effects appear in the electromagnetic response. 
Moreover, our results shed light on the origin of the quantum phase transition 
between the commensurate CDW phase and the stripe phase in TMD. 
 We also investigate the collective modes that appear when combined with low lying energy bulk  plasmons.

The article is organized in the following manner: in section II we
 discuss the plasmons in layered nodal liquids and specifically in TMD; 
in section III we develop the semi-classical calculation in general grounds 
and apply it to layered compounds like TMD in section IV. We have added an 
appendix where the details of the calculation are presented.   

\section{Nodal liquid plasmons}

The field theory of nodal liquids \cite{balents} is built under the basic idea
that the nodes of a CDW order parameter lead to two distinct subsystems
which correspond to the two components of the Dirac fermion spinor $\Phi_\sigma^\dagger = 
(\psi_{+\,\sigma}^\dagger , \psi_{-\,\sigma}^\dagger)$, with $+$,$-$ indexing
respectively the fermionic particles and anti-particles (holes). 
The non-interacting low energy Hamiltonian that describes the elementary excitations inside  
the Dirac cone is:
\begin{equation}
\mathcal{H}_D = \sum_\sigma \int_{BZ} \frac{\mathrm{d}^3 k}{(2\pi)^3}
\Phi_\sigma^\dagger(\mathbf{k}) \,\hbar \left( v_{F}\sigma_x k_x + 
v_{\Delta}\sigma_y k_y \right)\Phi_\sigma(\mathbf{k})\,, \label{H1}
\end{equation} 
where $\mathbf{k}$ is the momentum, $\sigma_{x,y}$ are Pauli matrices that
act in the particle-anti-particle subspace, 
$v_F$ and $v_\Delta$ are the anisotropic velocities perpendicular and parallel
respectively to the Fermi surface, and $BZ$ is the first 
 Brillouin zone of size $2\pi/d$ along the $k_z$ axis ($d$ the inter-plane distance).
When the chemical potential, $\mu$, intercepts the Dirac point,
the Hamiltonian (\ref{H1}) leads to the zero order polarization function \cite{Vozmediano} 
at $T=0$, whose complex conjugate is given by:
\begin{equation}
\Pi^{0\,*}(\omega,\mathbf{q}) = - \frac{v_F}{8 d\,\hbar v_{\Delta}} \, 
 \frac{\bar{q}^2}{\sqrt{v_F^2\bar{q}^2 - \omega^2}} \label{P}\,,
\end{equation} 
where $\omega$ is the frequency,
 $\bar{\mathbf{q}} = \vec{q}_x  + (v_\Delta/ v_F)\vec{q}_y$ the anisotropic
in-plane momentum. In the absence of hopping between the planes or interactions
with the lattice, the collective excitations of the Dirac fermions are due to the 3D Coulomb 
interaction, $V_0$, between carriers in different planes: 
$V_0(q,k_z) = 2\pi\,d\, e^2/(\epsilon_0 q) S(q,k_z)$,
with  a structure factor $ S(q,k_z) = \sinh(q d)/[\cosh(q d) - \cos(k_z d)]$ identical to 
the layered electron gas (LEG) case \cite{Hawrylak} ($e$ is the electron
charge, and $\epsilon_0$ the dielectric constant). In the random phase approximation (RPA) 
the electric  susceptibility is given by: $\epsilon(\omega,\mathbf{q},k_z) = 1 - V_0(q,k_z)\,
\Pi^{0}(\omega,\mathbf{q})$. At the points where the susceptibility vanishes,
that is, $\epsilon(\omega_p({\bf q},k_z),\mathbf{q},k_z)=0$, one obtains the plasmon 
dispersion, $\omega_p({\bf q},k_z)$. From (\ref{P}), we notice that no plasmon modes
 are allowed for Dirac fermions.

The displacement of the chemical potential from the Dirac point generates a pocket around the nodes
which drastically changes this picture. The density of states becomes finite
at the Fermi 
surface with Fermi momentum $k_F^*$, and Fermi energy $E_F^* = \hbar k_F^* v_F$, 
giving rise to intra-band excitations in the cone. 
In this case we recover the optic plasmon $\omega_p(q) = \sqrt{\omega_0^2 (\bar{q}/q)^2 + (v_0
\bar{q})^2}$ for the $k_z = 0$ bulk mode, where $v_0$ is the plasmon speed, 
and the 2D acoustic mode $\omega_p(q) \propto \sqrt{\bar{q}^2/q}$ for the rest
of the plasmon band $0<k_z \leq \pi/d$. The intra-band contribution is satisfactorily
understood in the context of doped graphite \cite{Shung}
where for $v_F\bar{q} \leq \omega < v_F (2k_F^* - \bar{q})$ 
there is gap in the particle-hole continuum of the 
inter-band excitations (see fig.1) defined by the imaginary part of (\ref{P}). 
The optical plasmon is free of Landau damping in
the long wavelength limit and its energy is of the order of 
$E_F^*$.  Notice that in the anisotropic case ($v_F \neq v_{\Delta}$) 
the gap in the plasmon spectrum depends on the direction
around the pocket Fermi surface. 

\begin{figure}[h!]
{\centering \resizebox*{2.4 in}{!}{\includegraphics{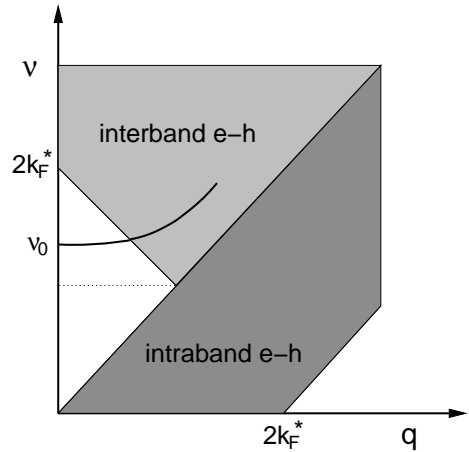}} \par}
\caption{{\small Schematic drawing of the $k_z = 0$ bulk plasmon in an isotropic pocket 
with Fermi momentum $k_F^*$. $\nu = \omega/v_F$ and $q$ is the
in-plane transfer momentum. The shaded areas correspond to the 
particle-hole continuum due to intra-band and inter-band excitations. }}
\label{excitation_spectrum}
\end{figure}

If we also add the piezoelectric electron-phonon coupling \cite{Neto} 
\begin{eqnarray}
\mathcal{H}_{EP} = \gamma  \sum_\sigma \int \mathrm{d}^3 x\,
\phi(\mathbf{x})\,\Phi_\sigma^\dagger(\mathbf{x}) \Phi_\sigma(\mathbf{x})\,,
\label{HEP}
\end{eqnarray}
due to acoustic phonons (with phonon field $\phi(\mathbf{x})$) 
with energy $\hbar \omega_{\mathbf{q}}$ into the pocket excitations, 
the RPA electric susceptibility acquires a correction:
$\epsilon(\omega,\mathbf{q},k_z) = 1 - [ V_0(q,k_z) + 
(\gamma^2/\hbar) D^0(\omega,\mathbf{q})] \Pi^{0}(\omega,\mathbf{q},\mu)
$ where \cite{fetter}
$$
D^0(\omega,\mathbf{q}) = \frac{\hbar \omega_{\mathbf{q}}^2 }{\omega^2 - (\omega_{\mathbf{q}} - i \eta)^2}\,,
$$ is the phonon propagator 
that affects very little the optical bulk plasmon in the  $q \rightarrow 0$ limit.
 
Besides doping, coherent inter-layer hopping is known to drive a dimensional crossover 
in direction to a 3D system, leading as well to a pocket formation in a nodal
liquid \cite{hopping}. The energy of the pocket is of the same order of the inter-layer hopping energy, 
which in 
TMD NbSe$_2$ and NbS$_2$ (we will drop the 2H prefix from now on) were calculated  in  $\sim$ 0.1 eV
 \cite{Doran}. We notice that inter-layer interactions were not taken into account into 
those band calculations  and that this energy could be considerably smaller. More experimental studies are required to investigate the nature 
of the low energy collective modes in these materials. However, we can consider the plasmon
modes in real materials to have a finite energy, like in ordinary metals. Nevertheless,
the plasmon frequency will be smaller than in ordinary metals because of the low
density of states in the system.

\section{Semi-classical dynamics}

We consider the problem of Dirac fermions described by (\ref{H1}) coupled to
lattice vibrations described by the Hamiltonian
\begin{eqnarray}
H_{PH} = \sum_n \frac{{\bf P}_n^2}{2} + \frac{1}{2} \sum_{n,m} K_{nm}
\left({\bf R}_n-{\bf R}_{m}\right)^2
\label{HPH}
\end{eqnarray}
where $n$ labels the lattice site $R^i_n$, $P^{i}_n$ is the lattice momentum
operator with $i=x,y,z$ (these operators are canonically conjugated, namely, 
$[P^i_n,R^j_m]=i \hbar \delta_{ij} \delta_{nm}$), $K_{nm}$ is the coupling
matrix. In what follows we will also consider the coupling
of the Dirac fermions and lattice degrees of freedom to a {\it classical} electromagnetic field
described by a vector potential field, ${\bf A}({\bf r})$, and electromagnetic
field tensor, $F^{\mu \nu} = \partial^{\mu} A^{\nu} - \partial^{\nu} A^{\mu}$
($\partial^{\mu} = \partial/\partial x_{\mu}$ with $\mu,\nu=0,i$)
with electromagnetic energy density:
\begin{eqnarray}
E_{el} = - \frac{1}{16 \pi} F_{\mu\nu} F^{\mu\nu}
\end{eqnarray}
where $F_{0j} = - E_j$ is the electric field strength (we use the metric 
$g_{\mu,\nu}=(1,-1,-1,-1)$) \cite{jackson}. In a piezoelectric the electric
field couples to the lattice distortion, $X^i_n =R^i_n-R^i_{n+1}$, 
via the piezoelectric tensor, $\Delta^i_{\,\,j}$, by:
\begin{eqnarray}
H_{P} = \sum_n \Delta^i_{\,\,j} E_i X^j_n
\label{HP}
\end{eqnarray}
where repeated indices are to be summed. 
The coupling between Dirac fermions and lattice vibrations
is given by (\ref{HEP}). The full problem can be written in path integral form
for the generating functional of the problem:
\begin{eqnarray}
Z = \int D{\bar \psi} D \psi \int D{\bf X} \exp\left\{\frac{i}{\hbar} \int dt \int d{\bf r} 
\mathcal{L}[\psi,{\bf X},A_{\mu}] \right\} \, ,
\end{eqnarray}
where $\mathcal{L}$ is the Lagrangian (real time) of the problem (${\bar \psi}$, $\psi$ are 
Grassmann variables).   

Let us consider first the case of the normal (non-superconducting) state. 
We are only interested
in the long wavelength physics of the electronic problem which, as argued in the previous
section, is described by the plasmon mode. The plasmon mode can be separated from the particle-hole
continuum using the formalism developed by Bohm and Pines \cite{bohm_pines} and the
Lagrangian of the problem reduces to:
\begin{eqnarray} 
\mathcal{L}_{N} &=& \mathcal{L}_{el} +  \mathcal{L}_{plasmon} +  
\mathcal{L}_{phonon} + 
\mathcal{L}_{piez}  \nonumber\\
        &=& - \frac{1}{16\pi} F_{\mu\nu}F^{\mu\nu} - \frac{1}{c} 
j_{N}^{\,\,\mu}A_{\mu}  +
     \frac{1}{2}  \Omega \left[ \omega_p^{-2}(\partial_{t}\rho_N)^2 - \rho_N^2 \right]\nonumber\\
&& + \frac{1}{2}\kappa\,[(\partial_{t} X^{i})^2  - v_{ph}^2 (\partial_i X^{i} )^2 ]
 + \Delta^{i}_{\,\,j}
X_{i}F^{0j} ,\qquad \label{L}
\end{eqnarray}
where $\Omega \approx 4 \pi/k_c^2$, with $k_c$  the 3D electron
screening wave vector (for an isotropic metal 
$k_c = \sqrt{3 \pi n e^2/(2 E_F)}$ where $n$ is the 3D electron density 
and $E_F$ the Fermi energy \cite{bohm_pines}), 
$\kappa$ is the lattice mass density,
$v_{ph}$ is the sound velocity, 
$j_{N}^{\,\,0}/c = \rho_{N}$ is the density of normal electrons and 
$j_{N}^{\,\,i}$ is the normal Ohmic current density (in our notation, 
$\partial^{0} = c^{-1}\partial_t $ is the time derivative
normalized by the speed of light $c$).

The collective properties of the electrons in the normal phase 
are the same of the electrons in the superconducting phase, since the opening 
of the superconducting gap does {\it not} affect the real part of the electronic polarization function,
 $\Pi(\mathbf{q})$, 
in the long wavelength limit \cite{Leggett}. In fact, within RPA one can easily show 
in the context of BCS theory that  $\mathrm{Re}\, \Pi(\mathbf{q})$ 
 remains essentially unchanged 
until the superconducting gap is
of the order of the Fermi energy, when the BCS pairing approximation becomes invalid
\cite{unpublished}.
In other words, for small $q$ the plasmon is not sensitive to the superconductor phase transition. 
Thus, normal and superconducting electrons screen electric fields exactly in the same way and   
therefore enter at equal footing in the Lagrangian. Thus, irrespective of the 
phase, normal or superconducting, $\rho_N$ in (\ref{L})
can be replaced by the {\it total} electron density $\rho=\rho_N + \rho_s$, 
where $\rho_s$ is the density of superconducting
electrons ($\rho_s=0$ in the normal state). Normal currents, however, are due to the quasi-particle 
excitations while the 
supercurrents, $j_{s}^{\,\,\mu}$, are due to the ground state of the condensate \cite{Bardeen}.
Thus, in the superconducting state the vector potential only couples to the normal currents.
At finite temperature, $T$, the number of normal and superconducting electrons is 
not  conserved. For simplicity, we have chosen to work at $T=0$, 
where all the electrons are in the condensate and the total current is $j^{\mu} =  j_{s}^{\,\,\mu}$.

In the superconducting phase we can introduce the superconducting order parameter, $\Psi$,
via a standard Hubbard-Stratanovich transformation and trace over the
electrons \cite{hub}. The generating functional reads:
\begin{eqnarray}
Z &=& \int D{\bar \Psi} D\Psi \int D{\bf X} \exp\left\{\frac{i}{\hbar} \int dt \int d{\bf r} 
\left(\mathcal{L}_N[\rho,{\bf X},{\bf A}] \right. \right.
\nonumber
\\
&+& \left. \left.
\mathcal{L}_{GL}[\Psi,{\bf A}]\right)\right\}
\label{Z}
\end{eqnarray}
where $\mathcal{L}_{GL}$ is the Ginzburg-Landau Lagrangian. 
In order to describe the macroscopic current-charge fluctuations we must also include a time 
dependence in the superconductor order parameter. This is not an obvious 
task but it can be done in two special limits: close to $T = 0$, and near the critical temperature 
\cite{Tsuneto}. In the later, the validity of the  
time-dependent Gorkov equations expansion in the gap, $\Delta_s$, require frequencies   
higher than the binding energy of the Cooper pairs, $\hbar \omega \gg \Delta_s$. In 
this limit the fluctuations have enough energy to break the Cooper pairs and 
convert them into single particle excitations, leading to a diffusive regime. 
As we mentioned above, we will work in the opposite limit $T = 0 $ where the hydrodynamic 
description at $\hbar \omega < \Delta_s$ is rigorously valid.
We introduce the low-temperature time-dependent Ginzburg-Landau Lagrangian 
\cite{Tsuneto,Parks},  
\begin{eqnarray}
\mathcal{L}_{GL} &=& - \alpha |\Psi|^2 - \beta |\Psi|^4 - \frac{1}{2m^\star} \left| 
\left( \frac{\hbar}{i} \nabla - 
\frac{e^\star}{c} \mathbf{A} \right)\Psi \right|^2 \nonumber \\
&& + \frac{1}{2
m^\star v_{s}^2} \left|\left( \frac{\hbar c}{i} \partial^{0} + e^\star \phi  
\right) \Psi  \right|^2\,.
\label{GL}
\end{eqnarray}
which is specially suitable for type II superconductors, where the penetration length is much 
larger than the coherence length (in (\ref{GL}) $\phi$ is the electrostatic 
potential and all the other symbols are standard \cite{tinkham}).
In the clean limit, $v_s$ is typically  of the order of the Fermi velocity $v_F$. 
Under the assumption that the fields vary much slower than the coherence length, we assume 
that the magnitude of the superconductor order parameter is 
constant and therefore all the fluctuations come from the superconducting
phase, $\varphi(t,\mathrm{x})$. Thus, the order parameter is written as:
\begin{eqnarray}  
\Psi(\mathrm{x}) = \Psi_{0}\,  \mathrm{e}^{i \varphi(t,\mathrm{x})} \, .
\end{eqnarray}
Despite the gauge invariance of (\ref{GL}),  we conveniently assume  
the transverse gauge, $\nabla \cdot \mathbf{A} = 0$, in what follows.  
Notice, however, that 
$\mathcal{L}_{N} + \mathcal{L}_{GL}$ is not complete because  
the fields and the supercurrents remain decoupled. This problem is solved by
assuming the validity of the non-homogeneous Maxwell equations. 
The Poisson equation is clearly a constraint between the density of 
superconducting electrons and the electrostatic potential,
$$
g(t, \mathbf{x}) \equiv  \nabla^2 \phi + 4 \pi \rho_{s} = 0 \,,
$$ 
and has to be enforced using a Lagrange multiplier $\Lambda(x^{\mu})$
in final Lagrangian.  

In the semi-classical regime, $\hbar \to 0$, 
the behavior of the fields comes from the minimal action principle
$$
\delta S = \delta \! \!\int \, \mathrm{d}^4 x \,\left( \mathcal{L}_{N} 
+\mathcal{L}_{GL}  + \Lambda\,g \right)  = \delta\!\! \int \, \mathrm{d}^4 x \,  
\mathcal{L} = 0\, ,
$$
with respect to the variables $A^{\mu}$, $\rho_{s}$,
$X^{i}$, $\varphi$ and $\Lambda$. 
By minimizing with respect to $A^{\mu}$ and $\Lambda$, we obtain 
the two non-homogeneous Maxwell's equations, 
\begin{equation}
 \nabla^{2} \phi  + 4 \pi \rho_{s} = 0 \label{phi1} 
\end{equation}
\begin{equation}
-  \square \mathbf{A} - \partial_{0} \nabla \phi + \frac{4 
\pi}{c} \mathbf{j}_{s} = 0 \label{A1}
\end{equation}
with the supercurrent $j_{s}^{\mu}$ given by
\begin{equation}
\rho_{s} = - \frac{e^\star \Psi_{0}^2}{m^\star v_{s}^2} \left(\hbar\,c\, 
\partial_{0} \varphi +  e^\star \phi \right)
-  \,\delta \,\nabla \cdot \mathbf{X} - \nabla^2 \Lambda \label{charge} 
\end{equation}
\begin{equation}
\mathbf{j}_{s} = \frac{e^\star \Psi_{0}^2}{m^\star}
 \left[\hbar\, \nabla \varphi  
- \frac{e^\star}{c} \mathbf{A}\right] +  \, c\,\delta \,\partial_{0}
\mathbf{X} \, .\label{j}
\end{equation}
For simplicity we will assume that $\Delta^{i}_{\,\,j} = \delta \delta_{ij}$ is a diagonal 
and isotropic tensor. The minimization with respect to the other fields 
completes the set of equations:
\begin{equation}
\frac{1}{4\pi}\,\Omega\,\left(\omega_p^{-2} \partial_t^2 \rho_{s} +  \rho_{s}\right)
 = \Lambda \label{rho}
\end{equation}
\begin{equation} 
 \nabla^2 \varphi
      -  \frac{c^2}{v_{s}^2}(\partial^{0})^2 \varphi -
\frac{e^\star c}{\hbar\,v_{s}^2}\, \partial^{0}\phi = 0 \label{varphid}
\end{equation}
\begin{equation}
 \kappa \left(  \partial_t^2 \mathbf{X} - v_{ph}^2 \nabla^2 \mathbf{X}\right)  -  
 \delta ( \nabla \phi + \partial^{0} \mathbf{A} )  = 0 \label{X} \,.
\end{equation}
Substituting  the supercurrent  
in the continuity equation, $\partial_{\mu} j_{s}^{\,\,\mu} = 0$, we find that
\begin{eqnarray}
c\,\partial_{0}\rho_{s} + \partial_{i} j_{s}^{\,\,i} 
= -\, c\,\partial_{0} \nabla^2 \Lambda = 0 \nonumber \,,
\end{eqnarray}
and therefore, in the absence of external currents $\Lambda$ is either a function of 
space or of time, but not of both. If $\Lambda$ depends on time, then $\rho$
is a stationary time dependent homogeneous (or non-periodic) distribution because of (\ref{rho}), 
what obviously violates the charge conservation. Therefore, 
we conclude that $\partial_0 \Lambda = 0$. 

The Lagrange multiplier defines two distinct classes of solutions for the charge density:
({\it i}) {\it screening-like} when $\Lambda(\mathbf{x}) \neq 0$, and ({\it
  ii}) {\it plasmon-like} when $\Lambda \equiv 0$. 
In the  first case, $ \rho(\omega,\mathbf{k}) \propto f(\mathbf{k}) 
\delta(\omega) $, where $f({\bf k})$ depends on the boundary conditions, has no dynamics
and describes the physical response to a boundary perturbation such as a 
static squeeze of the crystal or the introduction of a charge probe. In case
({\it ii}),  the electrons can oscillate freely with the plasmon frequency, 
allowing the existence of normal modes.

\section{Collective modes}

We consider the problem of a layered solid (such as the TMD) 
made by an infinite number of weakly interacting  
planes. In the continuum limit this problem becomes an effective 
3D model with spatial anisotropy in the direction perpendicular to the
planes, say the $z$ axis. It is convenient to define the
diagonal anisotropy tensor $\tau^{i}_{\,\,j} = (1,1,\tau )$,
where $\tau$ is the anisotropy parameter, (we introduce the notation 
$\tilde{\partial}^{i} \equiv \tau^{i}_{\,\,j}\partial^{j}$ and 
$\tilde{A}^{i} \equiv \tau^{i}_{\,\,j} A^{j}$) and rewrite 
the Ginzburg-Landau Lagrangian as a time-dependent Lawrence-Doniach model \cite{tinkham}: 
\begin{eqnarray}
\mathcal{L}_{GL} &=& - \alpha |\Psi|^2 - \beta |\Psi|^4 - \frac{1}{2m^\star} \left| 
\left( \frac{\hbar}{i} \tilde{\partial}^{i}\nonumber   - 
\frac{e^\star}{c} \tilde{\mathbf{A}} \right)\Psi \right|^2 \nonumber \\
&& + \frac{1}{2
m^\star v_{s}^2} \left|\left( \frac{\hbar c}{i} \partial^{0} + e^\star \phi  
\right) \Psi  \right|^2\,.
\label{GL2}
\end{eqnarray}
The model leads to the anisotropic version of equation 
(\ref{varphid}),
\begin{equation}
 \tilde{\nabla}\cdot \tilde{\nabla} \varphi
      -  \frac{c^2}{v_{s}^2}(\partial^{0})^2 \varphi -
\frac{e^\star c}{\hbar\,v_{s}^2}\, \partial^{0}\phi - 
\frac{e^\star}{\hbar\,c}\,\tilde{\nabla}\cdot \tilde{\mathbf{A}} = 0  \,,
\label{VarphiAnys}
\end{equation}  
and to the appearance of Josephson currents between the planes, 
represented by an anisotropic supercurrent:
\begin{equation}
\mathbf{j}_{s}(t,\mathbf{x}) = \frac{e^\star \Psi_{0}^2}{m^\star}
 \left[\hbar\, \tilde{\nabla} \varphi(t,\mathbf{x})  
- \frac{e^\star}{c} \tilde{\mathbf{A}}(t,\mathbf{x})\right] +  \, c\,\delta \,\partial_{0} 
\vec{X}(t,\mathbf{x}) \, .\label{jLD}
\end{equation}
From now on, we use an overhead symbol to represent the {\it in-plane} vectors $\vec{x}$, $\vec{q}$, $\vec{X}$,$\vec{\nabla}$ and etc. In our notation, we define $\vec{q}$ as the in-plane component of the momentum $\mathbf{k}$, such that $\mathbf{k} = (q,k_z)$ and $\tilde{\mathbf{k}} = (q,\tau k_z)$ . 

In highly anisotropic layered compounds the absence of piezoelectricity along the $z$
axis is justified by the weak distortions of the ions in this direction. 
The $z$-component of the sound velocity is also very small due to the weak elastic
coupling between planes. In the $v_{ph\,z}/v_{ph} \rightarrow 0$ limit, the phonon dispersion
is $\omega_{ph} = v_{ph}q $ and 
the in-plane phonon equation (\ref{X}) has a cylindrical symmetry 
\begin{eqnarray}
&& \kappa \left( - \omega^2  + \omega_{ph}^2 \right)\vec{X}(\omega, \mathbf{k}) \nonumber\\
  && \qquad\qquad -  \,i \,\delta \left[ \vec{q}\, \phi(\omega, 
\mathbf{k}) + \frac{\omega}{c} \vec{A}(\omega, \mathbf{k})\right]  = 0\,.\qquad 
\label{xAnys}
\end{eqnarray} 

Given the plasmon frequency, $\omega_p(\mathbf{k})$, 
the equation for the charge density in Fourier space becomes: 
\begin{equation}
\Omega
 \left[ - \omega^2 + \omega_{p}^2(\mathbf{k})\right] \rho_{s}(\omega,\mathbf{k}) =  
8\pi^2\,\omega_p^2(\mathbf{k})\,\Lambda(\mathbf{k})
\,\delta(\omega)\,.\label{rhoLD0}
\end{equation} 
The screening-like solution is:
\begin{equation}
\rho_{s}(\omega,\mathbf{k}) =  8\pi^2\,\frac{\Lambda(\mathbf{k})}{\Omega}
\,\delta(\omega)\,,\label{rhoLD}
\end{equation} 
for a non-zero $\Lambda(\mathbf{k})$. 
After a straightforward calculation, the simultaneous solution of equations 
(\ref{phi1})$-$(\ref{charge}), (\ref{VarphiAnys})$-$(\ref{xAnys}) and  (\ref{rhoLD})  
yields $\rho_{s}(\omega,\mathbf{k}) = \Lambda_{0}(\hat{\mathbf{k}})\,\delta\left(k^2 + k_{0}^2\right)$,
for some $\Lambda_0$ function, with the characteristic momentum 
\begin{equation}
k_{0}^2 = 4\pi \left( \frac{e^{\star\,2}\Psi_{0}^2}{m^\star\, v_{s}^2} - 
\frac{\delta^2}{\kappa\,v_{ph}^2} \label{k0}
\right)\,.
\end{equation}
The $k_0^2 > 0$ case leads to exponentially decaying solutions that describe the screening
induced by proper boundary conditions like the squeeze or shear of the crystal. 
If the piezoelectric coupling is larger than a critical value ($k_0^2 < 0$), the screening is suppressed
and a quasi-static charge modulation should be observed, in such way to 
minimize the elastic energy. This analysis is confirmed by  
introducing an external charge probe $Q$ at the origin:
we re-obtain the well known Thomas-Fermi result \cite{fetter},
\begin{equation}
\rho_{s}(\omega,\mathbf{k}) =  -\,2\pi\, \frac{Q\,k_{0}^2}{k^2 + k_{0}^2}\,\delta(\omega)\, ,
\end{equation}
which gives a screened potential for $k_0^2 > 0$ and recovers the metallic case when the cut-off of 
the in-plane band dispersion $s$ (of the order of inverse of the
in-plane lattice spacing) is taken to infinity. 

For a finite cut-off $s$ we have various cases. When $k_0 > s$, 
the screening is limited to the direction perpendicular to the planes.
When  $k_0^2 < 0$ and $|k_0| < s$, the system does not show any screening
since the potential decays like $\cos(|k_0| r)/r$ for large $r$. 
The last possible case, $k_0^2 < 0$ and $|k_0| > s$ is not 
physical in this theory and gives a purely imaginary response. 
We verify that the phase of the superconducting 
order parameter is free of fluctuations and satisfies the zero-field vortex equation
$(\vec{\nabla})^2 \varphi = 0$ in the whole $\Lambda \neq 0$ class.

The collective modes follow from a slightly different calculation. 
We start from the plasmon solution of (\ref{rhoLD0}),
\begin{equation}
\rho_{s}(\omega,\mathbf{k}) = \rho_0(\mathbf{k}) \,\delta(\omega - \omega_p(\mathbf{k}))\,,  
\end{equation}
where $\rho_0(\mathbf{k})$ is a  function which depends not only 
on the boundary conditions but also on the initial conditions driven by the perturbation.
In the non-relativistic 
limit $v_F/c \rightarrow 0 $, the field solutions in the momentum space are 
(the details of the calculation are given in the Appendix):
\begin{equation}
\rho_{s}(\omega,\mathbf{k}) = \rho(\hat{q}) \,\delta(q - q_{0})\,\delta(k_{z})
 \,\delta(\omega - \omega_p(\mathbf{k})) \, ,
\label{plasmon}\qquad
\end{equation}
\begin{equation}
\phi(\omega,\mathbf{k}) = 4\pi\, \frac{\rho_{s}(\omega,\mathbf{k})}{k^2} 
\, ,
\end{equation}
\begin{equation}
\varphi(\omega, \mathbf{k}) = 4\pi i \,\frac{\omega\,e^\star}{\hbar} 
\frac{\rho_s(\omega,\mathbf{k})}
{k^2} \frac{1}{\omega^2 - v_s^2 \tilde{k}^2} \, ,  
\label{VARPHIN}
\end{equation}
\begin{equation}
\vec{X}(\omega, \mathbf{k})
= 4\pi \frac{\delta}{i\,\kappa} \frac{\vec{q}}{k^2} 
\frac{\rho_{s}(\omega,\mathbf{k})}{(\omega^2 - \omega_{ph}^2)}\,\, ,
\end{equation}
\begin{equation}
\mathbf{A}(\omega, \mathbf{k}) =  4\pi\, \frac{\omega}{c} \vec{q}\, 
\frac{\rho_s(\omega,\mathbf{k})}{k^4}\,
\mathcal{D}(\omega, \mathbf{k}) \,,
\label{a}
\end{equation}
where we have labeled
\begin{equation}
\mathcal{D}(\omega, \mathbf{k}) =  1 - 4\pi \frac{e^{\star\,2}\Psi_{0}^2}{m^\star}
 \frac{1}{\omega^2 - v_s^2 \tilde{k}^2} + 4\pi \frac{\delta^2}{\kappa}\frac{1}
{\omega^2 - v_{ph}^2 q^2} \nonumber \,.
\end{equation}
Equation (\ref{plasmon}) represents the bulk plasmon mode $k_z = 0$, which prevails 
over the rest of the plasmon band. This is in agreement with  a general result valid for a 
stack of layers coupled by Coulomb interactions and by coherent hopping terms between adjacent planes.
The presence of inter-layer charge transfer induces
a dimensional crossover to a 3D system, which dominates the long wavelength spectrum. 
As we show in the end of the Appendix, this effect is conditioned to the  assumption that $\tau \neq 0$.
The inverse of the in-plane modulation scale $q_0$ is given by the equation 
 $\mathcal{D}(\omega_p,q_0) = 0$. 

Noting that at $T = 0$ the electronic density $n$ is twice the density of Cooper pairs $\Psi_0^2$, we realize that
the quantity  $4\pi e^{\star\,2}\Psi_0^2/m^\star$ can be conveniently written in the form of the expression that
gives the square of the plasma frequency, $\Omega_p^2  = 4\pi\,e^2 n/m $. 
In general grounds, $n$ is given by the sum rule \cite{tinkham}
\begin{equation}
\omega_p^2(k = 0)  = 
\frac{2}{\pi}\int_0^\infty \mathrm{d}\omega\, \omega\,\mathrm{Im} \,\epsilon(\omega,k)  = 
4\pi\, \frac{e^2 n}{m} f(\bar{q}/q)\label{SumRule}\,.
\end{equation}
where $\mathrm{Im}\,\epsilon$  is the imaginary  part of the electronic susceptibility, which is given in RPA by
$\mathrm{Im}\,\epsilon(\omega,\mathbf{k}) = - V_0(q,k_z) \mathrm{Im}\,\Pi^0(\omega,\vec{q},\mu)$ and 
$f(\bar{q}/q)$ is the anisotropy function due to the shape of the Fermi surface, with $f(\bar{q}/q) \equiv 1$
in the isotropic case. 
For  an anisotropic nodal liquid with
a small pocket we find that $f(\bar{q}/q) = \bar{q}^2/q^2$, as it may be easily checked by replacing  the 
leading intra-band polarization function 
\cite{Shung} 
$$
\mathrm{Im}\, \Pi^{0}(\omega,\vec{q},\mu) \stackrel{q/k_F^* \ll 1}{\longrightarrow} - \frac{k_F^*}{\pi \hbar v_\Delta d} 
\frac{\omega}{\sqrt{v_F^2\bar{q}^2 - \omega^2}}
$$
into (\ref{SumRule}), giving 
$$
4\pi \frac{e^{\star\,2}\Psi_{0}^2}{m^{\star}} = \omega_0^2   \,,
$$
with $\omega_0 = \sqrt{2k_F^* v_F^2 e^2/( d \epsilon_0^{\prime} \hbar v_\Delta)} $, where 
$\epsilon_0^\prime $ is the background susceptibility renormalized 
by the small inter-band terms \cite{Shung}. To simplify the analysis we consider 
the isotropic case only, where the rigid $\Psi_0$ approximation is more rigorous.
 Using the  $k_z = 0$  bulk plasmon mode $\omega_p^2 = \omega_0^2 + (v_0 q)^2$, we find that 
\begin{equation}
q_0 = \sqrt{ b - \sigma(b) \sqrt{b^2 - \frac{g \omega_0^2 }{(v_s^2 - v_0^2)
(v_{ph}^2 -v_0^2 )}}}\,,\label{K0}
\end{equation} 
where $\sigma$ is a sign function, $b = (g + \omega_0^2)/[2(v_{ph}^2 - v_0^2)] $ and 
\begin{eqnarray}
g = 4\pi \, \frac{\delta^2}{\kappa}
\label{geff}
\end{eqnarray}
is the effective piezoelectric coupling.  

Noting that $v_s > v_{ph}$, because the ions are much slower than the electrons, we identify two distinct cases, 
(1) $v_0 > v_s$ and  (2) $v_0 < v_s$.  The normal modes
are clearly absent in the regime (1) for any  value of  $g$. On the other hand, these modes are possible 
in the regime (2)  for non zero $g$. 
Naturally, they would not be physical for $g < (v_{ph}^2/v_s^2) \omega_0^2$, 
where the interactions are screened.

The clean limit of the superconductor gives exactly  
$v_s = v_F/\sqrt{3}$. In the particular case of an isotropic nodal liquid with 
a small pocket, the intra-band bulk plasmon disperses with $v_0 = (\sqrt{3}/2) v_F$ \cite{Shung} 
and therefore 
it corresponds to  the regime (1) mentioned above. This is also the case of the metal 
($v_0 = \sqrt{3/5} v_F$), 
and of doped graphite ($v_0 = (\sqrt{3}/2) v_F$).     
Therefore, in the particular case of TMD, 
 we should not observe 
any essential differences in the plasmon modes in comparison to the LEG (where $g = 0$) because of the piezoelectricity.

\subsection{Experimental results}

There is a large 
amount of experimental literature dealing with the observation of 
commensurate (CDW) or incommensurate charge order in TMD. Neutron and x-ray diffraction
measurements in  TaSe$_2$ and NbSe$_2$ reveal the existence of Bragg peaks at
incommensurate wave vectors $\mathbf{Q}_i = (1 - \delta_i)\mathbf{b}_i/3$,
where $\mathbf{b}_i$
($|\mathbf{b}_i| = 4\pi / (\sqrt{3}a)$) are the three reciprocal vectors with hexagonal symmetry, $a$
is the lattice spacing  
and $\delta_i \lesssim 0.02$ is the incommensurability
\cite{neutrons,stripes}. This state is called a triple CDW phase. 

In TaSe$_2$, the phase diagram temperature vs. pressure, $P$, is very rich 
\cite{McWhan} with three different phases: (1) a high temperature hexagonal incommensurate
CDW phase (HCDW) where the three ordering vectors have $\delta_i \neq 0$
($i=1,2,3$); (2) an
incommensurate stripe phase where one ordering vector is incommensurate, say
$\delta_1 \neq0$, but the other two are commensurate, $\delta_{2,3} = 0$;
(3) a commensurate CDW phase (CCDW) where $\delta_i = 0$ ($i=1,2,3$). 
The transition from the undistorted phase (normal) to HCDW occurs at a
pressure independent temperature, 
$T_{N-H} \approx 120 K$, up to pressures of $4.5$ GPa; the transition between
HCDW to stripe phase $T_{H-S} \approx 110 K$ is also roughly pressure independent; 
the transition between stripe phase and CCDW, $T_{S-C}(P)$, is highly pressure dependent
and vanishes at $P=P_c \approx 1.8$ GPa. Thus, there is a quantum phase
transition ($T=0$), as a function of pressure at $P=P_c$. By further
application of pressure there is a reentrant CCDW phase that has not been
fully studied and will be not discussed here. We will concentrate on the
nature of the quantum critical point (QCP) at $P=P_c$. 

The nature of the stripe phase has been discussed by McMillan \cite{mcmillan}
and others \cite{perbak} from the phenomenological point of view as a result
of the formation of topological defects of the complex CDW order parameter.
This stripe phase can be easily observed with electron microscopy \cite{Chen}.
The microscopic nature of this phase is still unknown but our results here
suggest that  it may have its origins on the piezoelectric coupling in
the solid. Indeed, we have found that there is a charge modulation in
these materials when $k_0^2<0$ (see eq.(26)). This modulation can be thought
as an incommensuration $\delta_i = 3 \sqrt{-k_0^2}/|{\bf b}_i|$. 
Equivalently, using definition
(\ref{geff}) there is a critical coupling constant $g_c$ so that an extra charge
modulation appears ($\delta_i >0$) when $k_0^2<0$ or 
\begin{eqnarray}
g > g_c = 4\pi\frac{v^2_{ph}}{v_s^2} \frac{e^{\star\,2}\Psi_{0}^2}{m^\star} \, .
\end{eqnarray}
Recall that TaSe$_2$ ($T_c \approx 0.1 K$) and NbSe$_2$ ($T_c
\approx 8.3 K$) are both superconductors \cite{Withers} whose $T_c$ increase
under application of pressure \cite{jap} and therefore $\Psi_0 \neq 0$ at
$T=0$. Notice that both the piezoelectric coupling, $\delta$, as well as
the sound velocity, $v_{ph}$, are also monotonically increasing functions
of pressure. If we assume that in TaSe$_2$ we have $g<g_c$ 
at $T=0$ and ambient pressure, then the
quantum phase transition can occur as a function of pressure as long as 
$g/v_{ph}^2$ is an increasing function of pressure. In NbSe$_2$, however,
the system is always incommensurate indicating that $g>g_c$ even at 
ambient pressure.
Since the $T=0$ phases seem to be directly connected with the stripe
phases that are observed at finite temperature we can immediately conclude
that the existence of these stripe phases have to do with the
piezoelectric coupling in these materials. Besides, recent experiments
 report a significant electrostatic modulation of $T_c$ in epitaxial bilayers composed by a
 HT$_c$ cuprate and a  polarized insulator deposited on it \cite{Matthey}. 
The connections
 between TMD and HT$_c$ are remarkable and we believe 
that this experimental result could be numerically calculated 
by relaxing the rigidity in the amplitude $\Psi_0$ of the superconductor order parameter. In addition, we
suggest that similar electrostatic devices combined with neutron scattering could test experimentally 
the role of the piezoelectricity in TMD  stripe phases.

Neutron scattering measurements have shown the softening of the
phonon optical mode
$\Sigma_1$ in NbSe$_2$ and TaSe$_2$ at the $\mathbf{Q}_i$ position for a wide range of temperatures
\cite{neutrons}. 
This behavior  was interpreted as a coupling of the optical phonons with the charge order.
Constant-$\mathbf{Q}$ scans at $\mathbf{Q}_i$ have shown that the phonon energy gap in both cases
 is of the order of 10  meV
\cite{neutrons}. 
 The coupling of the plasmons with the lattice vibrations 
in the present theory drives the long wavelength phonons to lock their frequency
with the frequency of the $k_z = 0$ bulk plasmon. 
In comparison to the metallic case where the plasmon
modes are rigid (in the sense that their wavelength has exactly the size of the system), 
we have found that the increase of the elastic energy due to the piezoelectric 
coupling may give rise to {\it elastic} plasmons which oscillate in resonance with the optical phonon
mode. Despite {\it not} observable in TMD, we believe that this effect would be the 
macroscopic manifestation of the plasmon-phonon resonance observed
 experimentally in these materials.

\section{Conclusions}

We have investigated the collective excitations of the electrons in a layered superconductor 
by a semi-classical calculation in the continuum limit of a highly anisotropic material. 
This procedure is analogous to the Lawrence-Doniach effective model for an 
infinite stack of layers.
Despite the evident interest in TMD, the calculation is sufficiently general and be could be applied to
any superconducting layered compound at zero temperature with broken lattice
inversion symmetry.   

We have demonstrated  from the 
electrodynamic point of view that superconductivity and piezoelectricity can coexist. 
Metallic screening is observed when the effective  piezoelectric coupling $g$ 
is smaller than a critical value $g_c$, with the Thomas-Fermi momentum reduced by the increase
of $g$. Above the critical coupling, the system is not screened and a long-range charge modulation is expected 
to appear as the response to 
a local quasi-static charge unbalance, which can be created by squeezing the crystal.
We have shown  that  piezoelectricity is possibly related to the mechanism behind the stripe formation in TMD.
 Besides, we have investigated the existence 
of  zero temperature 
normal modes arising in the presence of
low energy bulk plasmons, which dominate the spectrum of collective excitations.  We
have also  conjectured that the $k_z = 0$ plasmon mode behind these excitations
comes from the contribution of the intra-band excitations of a pocket opened around the nodes
of TMD by low energy coherent hopping terms between adjacent planes. 
These pockets could be also generated by doping TMD with intercalating materials.

\begin{acknowledgments}
A.~H.~C.~N. thanks C.~Varma for stimulating this work. We thank A.~J.~Leggett
for many discussions on the problem of screening in metals and superconductors.
B.U. and G.G.C. thank E. Miranda and Y. Copelevich for many helpfull discussions. They
also acknowledge Funda\c{c}\~ao de Amparo \`a Pesquisa do 
Estado de S\~ao Paulo (FAPESP), Brazil, project number 00/06881-9, for financial support.
B.U. thanks the Dept. of Physics at Boston University for the hospitality.   
\end{acknowledgments}

\appendix*
\section{}

Here we derive in detail the calculation of equations (\ref{plasmon})$-$(\ref{a}). Starting
from the plasmon-like solution  
\begin{equation}
\rho_{s}(\omega,\mathbf{k}) = \rho_{0}(\mathbf{k})\,\delta(\omega - \omega_p(\mathbf{k}))\,,
\end{equation} 
for a general function $\rho_{0}(\mathbf{k})$ and replacing it in the 
the Poisson equation (\ref{phi1}), we get
\begin{equation}
\phi(\omega,\mathbf{k}) = 4\pi \frac{\rho_{0}(\mathbf{k})}{k^2}\,
\delta(\omega - \omega_p(\mathbf{k}))\,. \label{phiNM}
\end{equation}

Next we separate the $z$-component of (\ref{A1}) and (\ref{jLD}),
\begin{eqnarray}
&& \frac{\omega}{c}
 k_{z}\,\phi(\omega,\mathbf{k}) + \left(\frac{\omega^2}{c^2} - 
k^2   
 -  \frac{4\pi}{c^2}\,\tau\,\frac{e^{\star\,2}\Psi_{0}^2}{m^\star} \right)
A_{z}(\omega, \mathbf{k}) \nonumber \\ 
&&  \qquad + \frac{4\pi}{c}\frac{e^\star\Psi_{0}^2}{m^\star} i \, \tau\,k_{z} \, \hbar\, 
\varphi(\omega,\mathbf{k}) = 0\, 
,\nonumber
\end{eqnarray}
and the $\varphi$ equation (\ref{VarphiAnys})
\begin{eqnarray}
&&\left(- \tilde{k}^2 + \frac{\omega^2}{v_{s}^2}  \right)\varphi(\omega, \mathbf{k})\nonumber\\
&& \qquad
- i\frac{e^\star}{\hbar\,c} \left(\frac{c\,\omega}{v_{s}^2}\phi(\omega, \mathbf{k}) +
\tilde{\mathbf{k}} \cdot \tilde{\mathbf{A}}(\omega,\mathbf{k}) \right) = 0\nonumber \, .
\end{eqnarray}
Noting that $\tilde{\mathbf{k}} \cdot \tilde{\mathbf{A}} =  k_{z}(\tau^2 - 1)A_{z} $,
for non-identically zero $\rho_{0}(\mathbf{k})$ we find
\begin{eqnarray}
&&\!\!\!\!\!\!\!\!\!\! A_{z}(\omega,\mathbf{k}) = -\frac{4 \pi}{c} \,\frac{\omega\,k_{z}}{k^2}\,
 \frac{\rho_{0}(\mathbf{k})}{\mathcal{F}(\omega,\mathbf{k} )}\,\delta(\omega - \omega_p(\mathbf{k}))
\nonumber \\
&&\!\!\!\!\!\! \times \left[ 1
- 4\pi \, \frac{e^{\star\,2}\Psi_{0}^2}{m^\star v_{s}^2 }
  \frac{\tau\,}{\mathcal{G}(\omega,\mathbf{k})} 
    \left( 
 1 -  \frac{v_{s}^2}{c^2} \frac{(\tau^2 - 1)k_{z}^2}{\mathcal{F}(\omega,\mathbf{k})}\right) 
  \right] \,. \label{AZ3} 
\end{eqnarray}
and
\begin{eqnarray}
\!\!\!\!\!\!\!\!\!\!\varphi(\omega, \mathbf{k}) &=& 4\pi i \,\frac{\omega\,e^\star}{\hbar\,v_{s}^2} 
\frac{\rho_{0}(\mathbf{k})}
{k^2} \frac{1}{\mathcal{G}(\omega, \mathbf{k})}\, \delta(\omega - \omega_p(\mathbf{k})) \nonumber\\ 
&&  \times \left[1 -  (\tau^2 - 1)
\frac{v_{s}^2}{c^2} \frac{k_{z}^2}{\mathcal{F}(\omega,\mathbf{k})}\right]
  \,, \label{VARPHINN}
\end{eqnarray}
where we have defined:
\begin{eqnarray}
\mathcal{F}(\omega,\mathbf{k}) &=& \frac{\omega^2}{c^2} - k^2   
 -  \frac{4\pi}{c^2}\,\tau\,\frac{e^{\star\,2}\Psi_{0}^2}{m^\star}\nonumber 
\end{eqnarray} 
and
\begin{eqnarray}
\mathcal{G}(\omega, \mathbf{k}) &=&  \frac{\omega^2}{v_{s}^2} - \tilde{k}^2  - 4\pi\,   
\frac{e^{\star\,2}\Psi_{0}^2}{m^\star c^2} 
\frac{ k_{z}^2 }{\mathcal{F}(\omega,\mathbf{k})} 
\,\tau \,(\tau^2 - 1)\, .\nonumber
\end{eqnarray}

The remaining results are derived from the phonon equation (\ref{xAnys}) 
$$
\vec{X}(\omega, \mathbf{k}) =  - i\,
 \frac{\delta}{\kappa}\frac{1}{\omega^2 - \omega_{ph}^2}
 \left(\vec{q}  \,
\phi(\omega, \mathbf{k})  + \frac{\omega}{c}
\vec{A}(\omega, \mathbf{k}) \right)
$$
and from  the combination of the in-plane components of (\ref{A1}) and (\ref{jLD}),
\begin{eqnarray}
 && \left(\frac{\omega^2}{c^2} - 
k^2  -  
\frac{4\pi}{c^2}\frac{e^{\star\,2}\Psi_{0}^2}{m^\star}\right)
\vec{A}(\omega, \mathbf{k})  + 
 \frac{\omega}{c}
 \vec{q}\,\phi(\omega,\mathbf{k})\qquad
\nonumber\\
&& \qquad
+ \frac{4\pi}{c}\frac{e^\star\Psi_{0}^2}{m^\star} i \,\vec{q}\,\hbar\, 
\varphi(\omega,\mathbf{k})  + \frac{4\pi}{c}i\,\omega\,\delta\,
\vec{X}(\omega,\mathbf{k}) = 0 \,. \nonumber
\end{eqnarray}
After a straightforward calculation, we encounter that
\begin{eqnarray}
\vec{A}(\omega, \mathbf{k}) &=& - \frac{4\pi}{c}\omega \vec{q}\, 
\frac{\rho_{0}(\mathbf{k})}{k^2}\,
\frac{\mathcal{D}(\omega, \mathbf{k})}{\mathcal{E}(\omega, \mathbf{k})}\,
\delta(\omega - \omega_p(\mathbf{k}))\nonumber\\
&&\label{AK}
\end{eqnarray}
\begin{eqnarray}
\vec{X}(\omega, \mathbf{k})
&=& - 4\pi i \frac{\delta}{\kappa} \frac{\vec{q}}{k^2} 
\frac{\rho_{0}(\mathbf{k})}{(\omega^2 - \omega_{ph}^2)}\, \delta(\omega - \omega_p(\mathbf{k}))
\nonumber  \\
 && \qquad \times \left[ 
1 -  \frac{\omega^2}{c^2} \frac{\mathcal{D}(\omega, \mathbf{k})}
{\mathcal{E}(\omega, \mathbf{k})}\right] \label{XA}\, .
\end{eqnarray}
where we have labeled
\begin{eqnarray}
\mathcal{D}(\omega, \mathbf{k}) &=&  1 + 4\pi \frac{\delta^2}{\kappa}\frac{1}
{\omega^2 - \omega_{ph}^2} - 4\pi \frac{e^{\star\,2}\Psi_{0}^2}{m^\star v_{s}^2}
 \frac{1}{\mathcal{G}(\omega, \mathbf{k})}\nonumber \\
&& \qquad \times \left[1 - (\tau^2 - 1)\,\frac{v_{s}^2}{c^2}
\frac{k_{z}^2}{\mathcal{F}(\omega, \mathbf{k})} \right] \nonumber
\end{eqnarray}
\begin{equation}
\mathcal{E}(\omega, \mathbf{k}) = \frac{\omega^2}{c^2} - k^2   
 -  \frac{4\pi}{c^2}\frac{e^{\star\,2}\Psi_{0}^2}{m^\star}\nonumber
 + 
\frac{4\pi}{c^2}\frac{(\delta\,\omega)^2}{\kappa\,(\omega^2 - \omega_{ph}^2)} \label{E}\, .
\end{equation}

Next, we substitute these results in the superconducting charge density definition (\ref{charge}) 
\begin{eqnarray}
&&\rho_{0}(\omega,\mathbf{k}) +  \frac{e^\star \Psi_{0}^2}{m^\star v_{s}^2}
\left[\hbar\,i\,\omega\,\varphi(\omega,\mathbf{k}) + e^\star
 \phi(\omega,\mathbf{k})  \right]\nonumber\\
&&\qquad\qquad + i\,\delta\,\vec{q}\cdot \vec{X}(\omega, \mathbf{k}) = 0\,,
 \nonumber
\end{eqnarray}
what yields
\begin{eqnarray}
&&  \rho_{0}(\mathbf{k})\,\delta(\omega - \omega_p(\mathbf{k})) \left\{ 1 -
 \frac{4\pi}{k^2} \frac{e^{\star\,2} \Psi_{0}^2}{m^\star v_{s}^2 }
\left[ \frac{\omega^2}{v_{s}^2}\,\frac{1}{ \mathcal{G}(\omega,\mathbf{k})}\,
 \nonumber \right.\right. \\
&& \qquad \times\left.  \left(1 -  (\tau^2 - 1)\frac{v_{s}^2}{c^2}
\frac{k_{z}^2}{\mathcal{F}(\omega,\mathbf{k})} \right)- 1\right]
 \nonumber \\
&& \qquad + \, \left. \, 4\pi \frac{\delta^2 }{\kappa} \frac{1}{k^2} 
\frac{q^2}{(\omega^2 - \omega_{ph}^2)} \left[ 
1 -  \frac{\omega^2}{c^2} \frac{\mathcal{D}(\omega, \mathbf{k})}{\mathcal{E}(\omega, 
\mathbf{k})}\right] \right\} \nonumber\\
&& =  0 \label{Eq2}
\end{eqnarray}

Applying  the transverse gauge to (\ref{AZ3}) and (\ref{AK}),
\begin{eqnarray}
\mathbf{k}\cdot \mathbf{A}
&=& - \frac{4 \pi}{c}\frac{\omega}{k^2}\, \rho_{0}(\mathbf{k})\,\delta(\omega - \omega_p(\mathbf{k}))
\nonumber \\
&&  \times \left\{\frac{k_{z}^2}{
\mathcal{F}(\omega,\mathbf{k})}\left[ 1 - 4\pi \frac{e^{\star\,2} \Psi^2}{m^\star v_{s}^2} 
\frac{\tau}{\mathcal{G(\omega,\mathbf{k})}} \right.\right. \nonumber\\
&& \left.\left. \times \left( 
 1 - (\tau^2 - 1) \frac{v_{s}^2}{c^2} \frac{k_{z}^2}{\mathcal{F}(\omega,\mathbf{k})}\right)
  \right]  +  q^2 \frac{ \mathcal{D}(\omega,\mathbf{k})}
{\mathcal{E}(\omega,\mathbf{k})}   \right\} \nonumber \\
&=& 0 \label{GAUGE}
\end{eqnarray} 

This way, we conclude that $\rho_{0}(\mathbf{k})$ is on the form
$$
\rho_{0}(\mathbf{k}) = \rho(\hat{q}) \,\delta(q - q_{0})\,\delta(k_{z} - k_{z\,0}) \, , 
$$ 
where $q_{0}$ and $k_{z\,0}$ are the zeroes   
of (\ref{GAUGE}) and (\ref{Eq2}).
In the non-relativistic limit $v_F/c \rightarrow 0$, 
 $\mathcal{F}(\omega_p, \mathbf{k}) \sim 
\mathcal{E}(\omega_p, \mathbf{k}) \rightarrow  - k^2$ and 
\begin{equation}
\mathcal{G}(\omega, \mathbf{k}) \rightarrow \frac{\omega^2}{v_{s}^2} - \tilde{k}^2
\nonumber
\end{equation}
\begin{equation}
\mathcal{D}(\omega, \mathbf{k}) \rightarrow 1  - 4\pi \frac{e^{\star\,2} \Psi_{0}^2}{m^\star} 
\frac{1}{ \omega^2 - v_{s}^2 \tilde{k}^2 } + 4\pi \frac{\delta^2}{\kappa}\frac{1}
{\omega^2 - \omega_{ph}^2} \,.\nonumber
\end{equation}
In this limit, (\ref{GAUGE}) and (\ref{Eq2}) simplify respectively to
\begin{eqnarray}
&& \left( k^2  +  \, 4\pi \frac{\delta^2}{\kappa} 
\frac{ q^2}{\omega^2 - \omega_{ph}^2}  -
 4\pi  \frac{e^{\star\,2} \Psi_{0}^2}{m^\star  }
 \frac{q^2 + \tau\,k_{z}^2 }{\omega^2  - v_{s}^2\tilde{k}^2}
   \right)\nonumber \\
&& \qquad \times
 \rho_s(\omega,\mathbf{k})\,\delta(\omega - \omega_p(\mathbf{k})) = 0 
\end{eqnarray}
\begin{eqnarray}
&&\left( k^2  +  \, 4\pi \frac{\delta^2}{\kappa} 
\frac{q^2}{\omega^2 - \omega_{ph}^2}  -
 4\pi  \frac{e^{\star\,2} \Psi_{0}^2}{m^\star  }
 \frac{\tilde{k}^2}{\omega^2  - v_{s}^2\tilde{k}^2}
   \right)\nonumber \\
&& \qquad \times \rho_s(\omega,\mathbf{k})\delta(\omega - \omega_p(\mathbf{k})) = 0\, .
\end{eqnarray} 
Comparing both and recalling that $\tilde{k}^2 = q^2 + \tau^2 k_{z}^2$, then 
$ \tau\, (1 - \tau) \,k_{z}^2 = 0$. For  $\tau \neq 0,1$, we immediately see that $k_{z\,0} = 0$. 
Substituting this result in one of the expressions above and integrating in $\omega$, we find
that $\mathcal{D}(\omega_p, q_0) = 0$.

\end{document}